\documentclass[twocolumn]{revtex4-1}

\usepackage{graphicx}
\usepackage{amsfonts}
\usepackage{amsmath}
\usepackage{hyperref}

\newcommand{\bra}[1]{{\langle}#1|}
\newcommand{\ket}[1]{|#1{\rangle}}
\newcommand{\expect}[1]{\langle {#1} \rangle}

\begin{document}

\title{Optimized geometries for future generation optical lattice clocks}

\author{S. Kr\"amer}
\affiliation{Institute for Theoretical Physics, Universit\"at Innsbruck, Technikerstra{\ss}e 21/3, 6020 Innsbruck, Austria}
\author{L. Ostermann}
\affiliation{Institute for Theoretical Physics, Universit\"at Innsbruck, Technikerstra{\ss}e 21/3, 6020 Innsbruck, Austria}
\author{H. Ritsch}
\affiliation{Institute for Theoretical Physics, Universit\"at Innsbruck, Technikerstra{\ss}e 21/3, 6020 Innsbruck, Austria}

\begin{abstract}
Atoms deeply trapped in magic wavelength optical lattices provide a Doppler- and collision-free dense ensemble of quantum emitters ideal for high precision spectroscopy. Thus, they are the basis of some of the best optical clock setups to date. However, despite their minute optical dipole moments the inherent long range dipole-dipole interactions in such lattices generate line shifts, dephasing and modified decay. We show that in a perfectly filled lattice these effects are resonantly enhanced depending on lattice constant, lattice geometry and excitation scheme inducing clock shifts of many atomic linewidths and reducing measurement precision via superradiance. However, under optimal conditions collective effects can be exploited to yield zero effective shifts and prolong dipole lifetimes beyond the single atom decay. In particular we identify 2D hexagonal or square lattices with lattice constants below the optical wavelength as most promising configurations for an accuracy and precision well below the independent ensemble limit. This geometry should also be an ideal basis for related applications such as superradiant lasers, precision magnetometry or long lived quantum memories.
\end{abstract}

\maketitle

\noindent Since the turn of the century the technology of manipulating and controlling ultracold atoms with lasers has seen breathtaking advances~\cite{bloch2012quantum, lewenstein2007ultracold, micheli2006toolbox}. Following the seminal first demonstration of a quantum phase transition in an optical lattice~\cite{greiner2002quantum}, nowadays the so-called Mott insulator state can be prepared routinely~\cite{bloch2005ultracold, bakr2009quantum}. Experiments with photo-associated ultracold molecules have reached a comparable control~\cite{ospelkaus2006ultracold, carr2009cold, danzl2010ultracold, yan2013observation} and coherent interactions between the atoms at neighboring sites can be tailored~\cite{duan2003controlling}.

For some of the world's best optical clocks~\cite{martin2013quantum, ushijima2014cryogenic, zhang2014spectroscopic}, atoms with a long-lived clock transition are prepared in an optical lattice using a differential light shift free (magic) trapping wavelength~\cite{takamoto2005optical, ludlow2008sr}. In principle, this provides for a Doppler and collision free dense ensemble with negligible inhomogeneous broadening. However, when excited optically emitters will nevertheless interact via long range resonant dipole-dipole coupling~\cite{ficek1986cooperative}.

Here we show that at sufficient densities the dipole interaction strength surpasses the excited state lifetime and collective excitations analogous to excitons appear~\cite{zoubi2013excitons}. For polar molecules in optical lattices they even dominate the dynamics~\cite{pollet2010supersolid} and allow for studying generic phenomena of solid state physics~\cite{bloch2012quantum}. For clock transitions the extremely tiny dipole moment keeps these interactions small in absolute magnitude. Still, the exciton's effective transition frequencies and their spontaneous decay is governed by dipole-dipole interaction~\cite{ostermann2012cascaded} deviating from the bare atom case. This limits accuracy and precision of corresponding clock setups.
\begin{figure}[h]
  \center
  \includegraphics{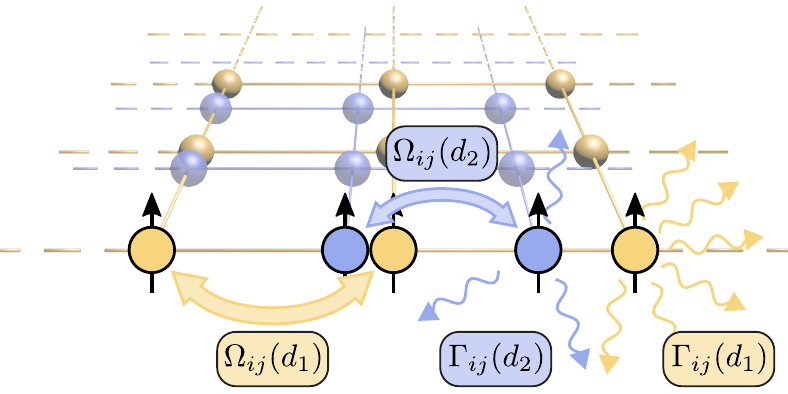}
  \caption{(Color online) Scheme of a 2D optical lattice filled with clock atoms interacting via dipole-dipole energy exchange $\Omega_{ij}$ and a collectively modified spontaneous emission $\Gamma_{ij}$ at two different lattice constants shown in blue and yellow. In a mean field treatment with translation invariance the sum over all interaction terms yields two effective couplings $\Omega^\mathrm{eff}$ and $\Gamma^\mathrm{eff}$ only, which govern the approximate system dynamics.}
  \label{fig:system}
\end{figure}
In an idealized Ramsey sequence for a clock setup, the first laser pulse creates a product state of all atoms prepared in a $50\%$ superposition of ground and excited state with equal phase and all dipoles aligned in parallel. This state features the maximally possible dipole moment and typically exhibits superradiance. Even a tiny single particle spontaneous emission rate can be that strongly enhanced, that collective decay becomes a dominant factor limiting measurement time and precision~\cite{ostermann2013protected}. In current setups based on 1D lattices with low filling this perturbation is often negligible compared to other noise like collisions, black body shifts or reference cavity fluctuations. However, in lattices with unit filling, dipole-dipole interaction shifts are larger than the atomic linewidth and constitute a significant inherent perturbation. Note that their absolute magnitude scales with the atomic dipole moment and thus strongly depends on the chosen transition.

As a key quantity to capture the collective modifications of the system dynamics, we use the decay and phase shift of the collective dipole generated by the first Ramsey pulse, which determines the contrast and shift of the central Ramsey fringes. Note, that due to the pairwise nature of dipole-dipole interactions a rephasing pulse cannot correct these errors. Here we ignore interaction induced perturbations during the Ramsey pulses, which introduce extra noise but could be reduced by very fast pulses or improved sequences~\cite{yudin2010hyper}.

For this we numerically solve the well established master equation for the atomic density matrix $\rho$ including optical dipole-dipole interaction obtained by tracing over the electromagnetic vacuum field~\cite{ficek1986cooperative,gardiner2004quantum,Lehmberg1970radiation},
\begin{equation}
  \dot{\rho} = i \left[ \rho, H \right] + \mathcal{L}[\rho].
  \label{eq:master-equation}
\end{equation}
As previously shown for small atom numbers ($N<12$) a numerical solution of the full master equation yields non-negligible shifts already~\cite{chang2004controlling, ostermann2012cascaded}. As the Hilbert space grows exponentially with atom number, the full equation cannot be solved for ensembles of a realistic size. Since for precision measurements we need to evaluate collective effects precisely, reliable and converging alternative numerical methods are required. For larger ensembles at low densities a cluster approach has produced first estimates of the scaling of the dephasing with the system's size and density~\cite{martin2013quantum}. Recently, important self synchronization effects through dipole coupling were studied in a very high density limit using simplifying assumptions for the coupling~\cite{zhu2014quantum}.

In this letter we present an extensive analysis of the collective dynamics for fully populated lattices of different geometries and sizes containing a large or even an infinite number of particles. Our primary goal is to estimate the magnitude of the dipole phase shift and collective decay as a function of lattice and excitation geometry. Besides a resonant enhancement of shifts, decay and dephasing, we find cases where collective effects lead to improvements of the maximally achievable measurement precision beyond the independent particle level by virtue of subradiant states. We concentrate on an idealized setup ignoring lattice shifts, thermal effects or the hopping of atoms.

Numerically we apply an enhanced mean field approach whose validity has been extensively tested in recent work~\cite{Kraemer2015Generalized}, which we can scale to realistic numbers up to $N \approx 10^5$ particles. If the particle distribution exhibits symmetries numbers up to even $10^{10}$ are possible, well approximating infinite systems in 1D and 2D. Its accuracy, however, breaks down at very close distances as it cannot capture higher order correlations. Similar deliberations for classical dipoles have recently been put forward~\cite{bettles2014cooperative}.

{\it Model --} We consider an ensemble of $N$ identical effective two-level atoms with transition frequency $\omega_0$ and inverse lifetime $\gamma$ at positions $r_i$ ($i=1..N$) interacting via optical dipole-dipole coupling described by the Hamiltonian~\cite{Lehmberg1970radiation,ficek1986cooperative}
\begin{equation}
  H = \sum_{ij;i \neq j} \Omega_{ij} (r_{ij}) \sigma_i^+ \sigma_j^-.
  \label{eq:hamiltonian}
\end{equation}
Here, $\sigma^\pm_i$ denotes the raising (lowering) operator of the $i$-th atom and $\Omega_{ij} = \frac{3}{4} \gamma G \left(k_0 r_{ij}\right)$ represents the energy exchange with $k_0 = \omega_0/c = 2\pi/\lambda_0$ and $r_{ij} = \left| r_i - r_j \right|$ being the distance between atoms $i$ and $j$. Collective spontaneous emission is accounted for by a Liouvillian of the form~\cite{ficek1986cooperative,gardiner2004quantum}
\begin{equation}
  \mathcal{L}[\rho] = \frac{1}{2} \sum_{i,j} \Gamma_{ij}(r_{ij})
                        (2\sigma_i^- \rho \sigma_j^+
                        - \sigma_i^+ \sigma_j^- \rho
                        - \rho \sigma_i^+ \sigma_j^-),
  \label{eq:lindblad}
\end{equation}
where the off-diagonal rates $\Gamma_{ij} = \frac{3}{2} \gamma F \left( k_0 r_{ij} \right)$ introduce super- and subradiant decay~\cite{Lehmberg1970radiation}. Explicitly we have
\begin{subequations}
\begin{align}
  F(\xi) &= \alpha \frac{\sin \xi}{\xi}
            + \beta \left(
                  \frac{\cos \xi}{\xi^2} - \frac{\sin \xi}{\xi^3}
            \right)
  \\
  G(\xi) &= -\alpha \frac{\cos \xi}{\xi} + \beta \left(
                \frac{\sin \xi}{\xi^2} + \frac{\cos \xi}{\xi^3}
            \right)
\end{align}
\end{subequations}
with $\alpha = 1 -\cos^2 \theta$ and $\beta = 1-3 \cos^2 \theta$, where $\theta$ represents the angle between the line connecting atoms $i$ and $j$ and the common atomic dipole orientation.

{\it Mean field approximation --} To study large particle numbers we derive the equations of motion for the expectation values of the Pauli operators for the $k$-th atom as detailed in the appendix~\ref{appendix:meanfield-equations}. Assuming a separable density operator and factorizing the two-particle correlations via $\expect{\sigma^\mu_i \sigma^\nu_j} \approx \expect{\sigma^\mu_i}\expect{\sigma^\nu_j}$ for $\mu,\,\nu \in \{x, y, z\}$ they transform to a closed set. As shown previously~\cite{Kraemer2015Generalized} these equations still capture the major part of the interaction up to a moderate interaction strength. For more accurate studies one may add second order corrections increasing the computational effort.

{\it Symmetric configurations --} For symmetric geometries with each atom initially in the same state and subject to the same effective interactions, the equations of motion for all particles become identical and read
\begin{subequations}
\begin{align}
  \expect{\dot{\sigma^x}} &=
      \Omega^{\mathrm{eff}}\expect{\sigma^y}\expect{\sigma^z}
      -\frac{1}{2} \Big(
          \gamma
        -\Gamma^{\mathrm{eff}}\expect{\sigma^z}
      \Big) \expect{\sigma^x},
  \\
  \expect{\dot{\sigma^y}} &=
      -\Omega^{\mathrm{eff}}\expect{\sigma^x}\expect{\sigma^z}
      -\frac{1}{2} \Big(
        \gamma
        -\Gamma^{\mathrm{eff}}\expect{\sigma^z}
      \Big) \expect{\sigma^y},
  \\
  \expect{\dot{\sigma^z}} &=
        -\gamma \big(1 + \expect{\sigma^z}\big)
        -\frac{1}{2} \Gamma^{\mathrm{eff}} \Big(\expect{\sigma^x}^2 + \expect{\sigma^y}^2\Big).
  \label{eq:meanfield-equations}
\end{align}
\end{subequations}
Hence instead of solving a huge set of coupled nonlinear equations we need to determine the effective couplings, i.e.
\begin{equation}
  \Omega^\mathrm{eff} = \sum_{j=2}^N \Omega_{1j}
  \qquad
  \Gamma^\mathrm{eff} = \sum_{j=2}^N \Gamma_{1j},
  \label{eq:effective-quantities}
\end{equation}
Of course, such a rigorous symmetry condition is fulfilled for very few atomic distributions only. Then, the essence of the interactions within the entire lattice is captured solely by two real numbers, the effective coupling $\Omega^{\mathrm{eff}}$ and the collective decay rate $\Gamma^{\mathrm{eff}}$. In a clock setup one seeks to minimize the energy shifts $\Omega^\mathrm{eff}$ and find configurations with a maximally negative $\Gamma^\mathrm{eff}$, minimizing decay and allowing for an as long as possible interrogation time.

\begin{figure}[h]
  \center
  \includegraphics{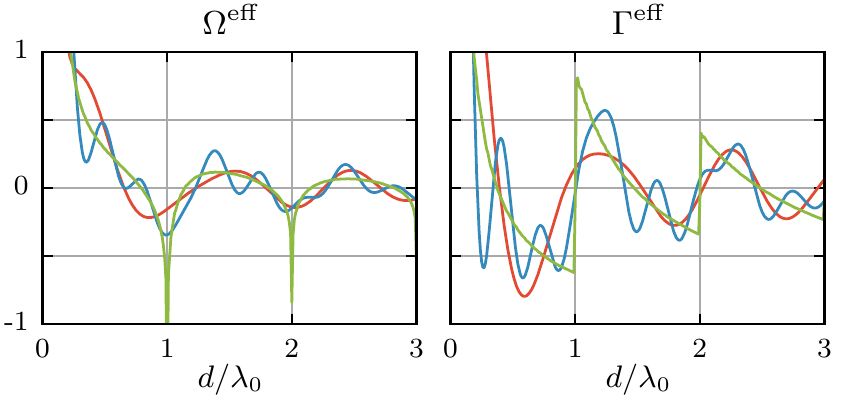}
  \caption{(Color online) Distance dependence of the effective dipole coupling $\Omega^\mathrm{eff}$ and $\Gamma^\mathrm{eff}$ for a square (red), a ten-sided (blue) and a 100000-sided (green) regular polygon. The fewer particles the closer the functions resemble the underlying couplings $\Omega_{ij}$ and $\Gamma_{ij}$. The divergences at integral $d/\lambda_0$ result from the $1/\xi$ terms in $F(\xi)$ and $G(\xi)$.}
  \label{fig:effectiveinteraction-polygons}
\end{figure}

{\it Finite systems --} Firstly, for finite symmetric configurations the effective quantities can be calculated easily. The most obvious symmetric structures are regular polygons. This might not be the most practical example but nicely displays the underlying physics. In Fig.~\ref{fig:effectiveinteraction-polygons} we compare the parameters for a square, a ten-sided and a 100000-sided polygon. The square shows a behavior quite similar to the underlying functions $F(\xi)$ and $G(\xi)$, while the two larger polygons exhibit strong size dependent variations, particularly at integral values of $d\lambda_0$ emerging from the accumulation of many $1/\xi$ contributions. Note that even with a relatively large atom spacing, cooperative collective effects are sizable and vary strongly with distance.
\begin{figure*}[t]
  \center
  \includegraphics{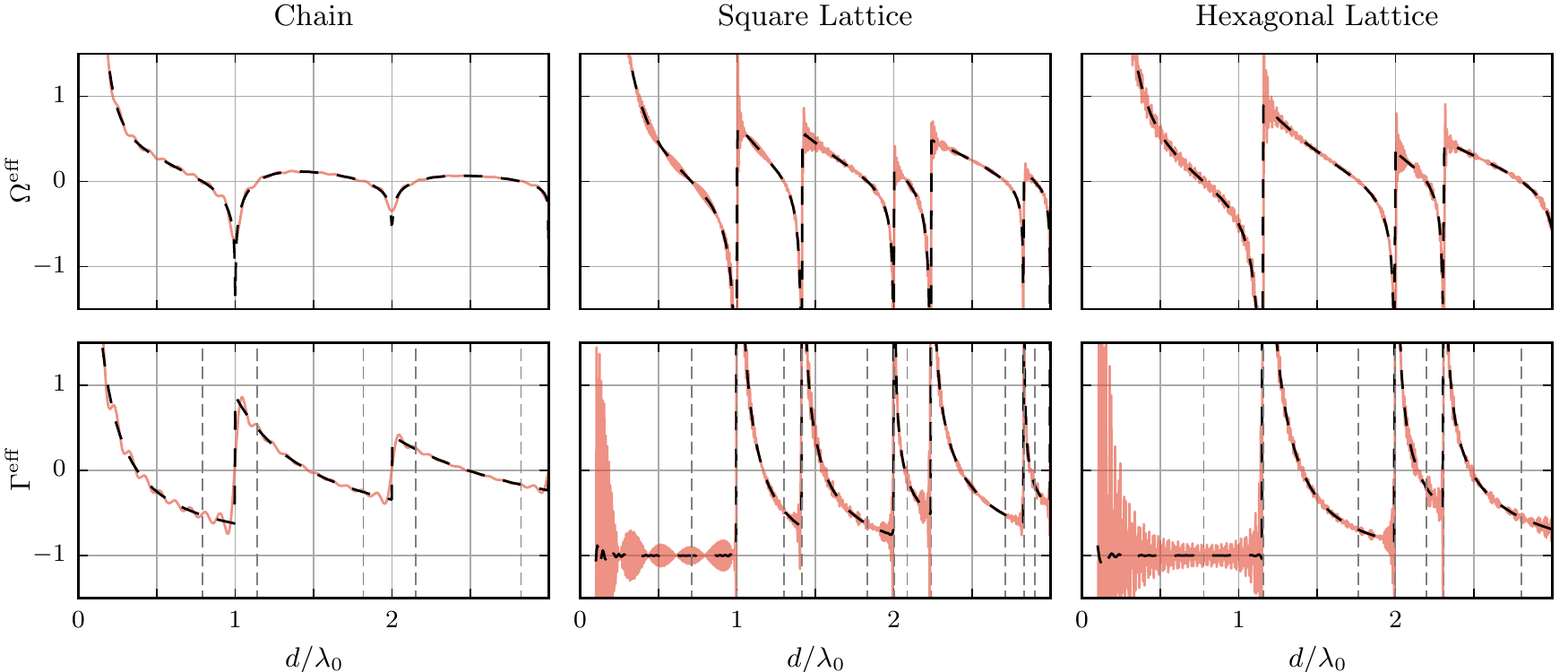}
  \caption{(Color online) Distance dependence of the effective quantities $\Omega^\mathrm{eff}$ and $\Gamma^\mathrm{eff}$ for an infinite equidistant chain, a square lattice and a hexagonal lattice (dashed black) compared to their not yet converged finite counterparts of 10, $4*10^4$ and $10^5$ particles respectively (solid red). Again, we find divergences at integral $d/\lambda_0$ owing to the $1/\xi$-terms in $F(\xi)$ and $G(\xi)$. In the 2D configurations $\Gamma^\mathrm{eff}$ plateaus at $-1$ for $d < \lambda_0$, suggesting that this parameter range will be the most favorable for clock setups as decay is strongly suppressed.}
  \label{fig:effectiveinteraction-infinitesystems}
\end{figure*}

{\it Infinite systems --} In practice, extended regular systems, i.e.\ large periodic lattices, are experimentally more relevant. Fig.~\ref{fig:effectiveinteraction-infinitesystems} depicts the effective couplings for an infinite chain, a square lattice and a hexagonal lattice. For comparison, we have overlaid the results for smaller atom numbers to demonstrate finite size effects. We observe stronger variations and again divergences at integral values of $d/\lambda_0$ . These manifest themselves in a much more pronounced way at huge atom numbers and therefore underpin the importance of properly treating long range interactions.

Note that for the two-dimensional square lattice and the hexagonal lattice $\Gamma^\mathrm{eff}$ exhibits a broad minimum for the effective decay close to $\Gamma^\mathrm{eff}=-1$ for $d < \lambda_0$, where atomic decay is strongly inhibited. This favors  such two-dimensional setups for lattice clocks as subradiant decay will dominate the system dynamics allowing for much longer Ramsey delay times and thus offering a higher overall precision~\cite{ostermann2013protected}. Similarly we can identify lattice constants with a zero effective shift increasing clock accuracy. Extending these calculations to three dimensional lattices, we find that the necessary atom numbers to obtain smooth converging behavior are beyond our current numerical capabilities. For particle numbers of about $10^{12}$ the resulting effective quantities still fluctuate strongly, predicting potential problems for such 3D clock setups. A demonstration of this effect can be found in the appendix~\ref{appendix:effective-quantities-cubic}.

\begin{figure}[ht]
  \center
  \includegraphics{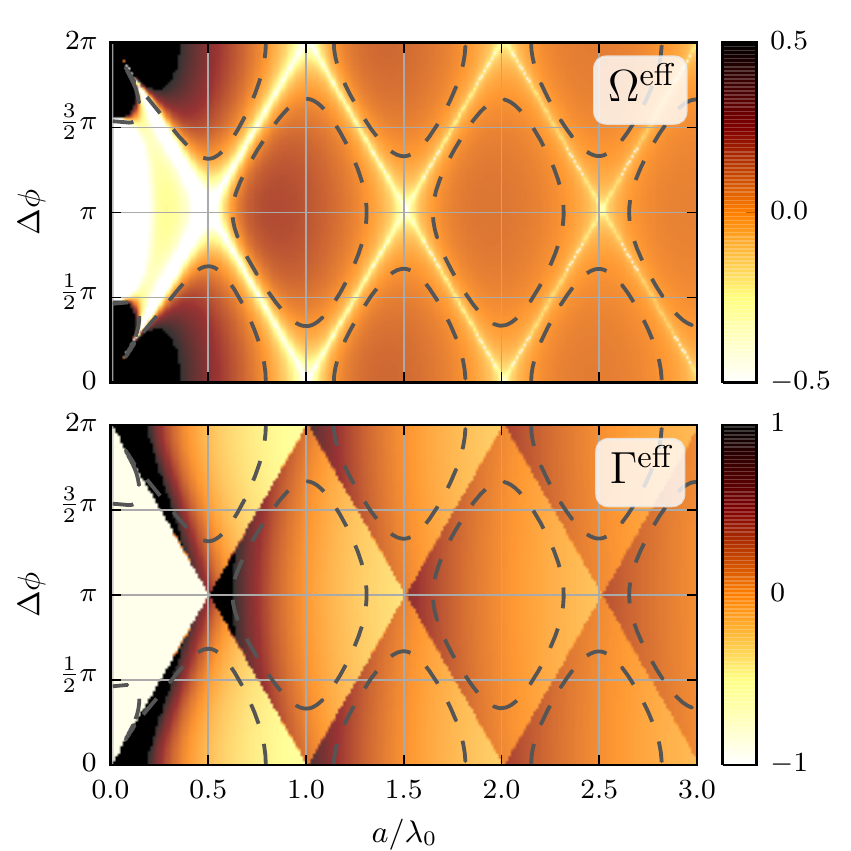}
  \caption{(Color online) Effective interactions $\Omega^\mathrm{eff}$ and $\Gamma^\mathrm{eff}$ for an infinite chain with spacing $a$ where the spins are initially prepared with phase shift $\Delta \phi$ between neighboring spins. The dashed lines indicates parameters with $\Omega^\mathrm{eff}=0$ optimal for an optical clock.}
  \label{fig:effectiveinteractions-phase-chain}
\end{figure}

\begin{figure}[ht]
  \center
  \includegraphics{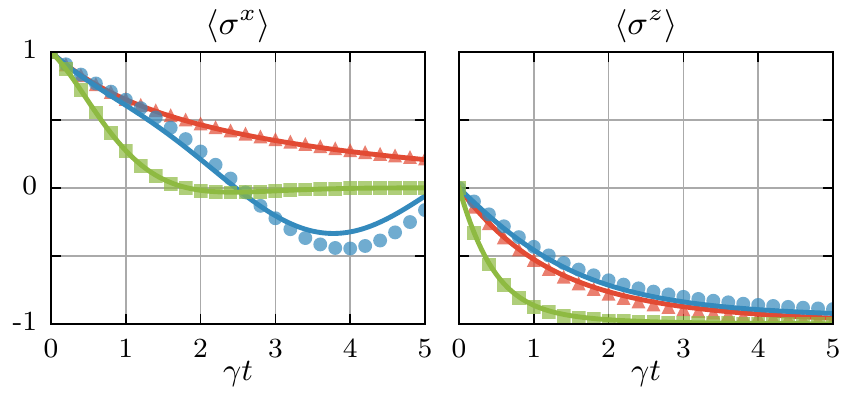}
  \caption{(Color online) Three different examples for the time evolution of the spin expectation values for a chain with spacing $d$ where initially all spins are prepared in a coherent superposition of ground and excited state with a phase shift of $\Delta \phi$. The parameters used are $d=0.792\lambda_0$ with $\Delta \phi=0$ (red triangles), where $\Omega^\mathrm{eff}=0$ and $\Gamma^\mathrm{eff}$ is nearly optimal, as well as $d=0.49\lambda_0$ (green squares) and $d=0.51\lambda_0$ (blue circles), both with $\Delta \phi=\pi$ which are close to a $\Gamma^\mathrm{eff}$ discontinuity. The solid lines correspond to a solution of a second order cumulant expansion model with 200 particles and demonstrate a very good agreement with the infinite mean field description.}
  \label{fig:timeevolution-chain}
\end{figure}

{\it Tailoring atomic excitations --} So far we assumed a phase-symmetric excitation of all atoms by the first Ramsey pulse. In a practical excitation scheme this corresponds to illumination at right angle. In general, however, the effective couplings $\Omega^\mathrm{eff}$ and $\Gamma^\mathrm{eff}$ will change, when we allow for a local phase shift imprinted on the atoms. In a $\pi/2$ Ramsey sequence~\cite{haroche2006exploring} the excitation phase appears on the excited state directly, i.e.
\begin{equation}
  \ket{\Psi} =
    \bigotimes_{j=1}^N \frac{1}{\sqrt{2}}
      \left(
        \ket{g} + e^{i \Delta \phi (j-1)} \ket{e}
      \right).
  \label{eq:ramsey-state}
\end{equation}
In our treatment we can exploit the system's symmetry and absorb this phase into the effective couplings (Appendix~\ref{appendix:meanfield-phase-equations}). For $\Delta \phi = 0$ we recover the above results. The closer the phase shift gets to $\Delta \phi = \pi$, however, the more half-integral values of $d\lambda_0$ yield minimal shifts and the maximally negative $\Gamma^\mathrm{eff}$ as seen in Fig.~\ref{fig:effectiveinteractions-phase-chain}. Since the emitted light has interfered constructively at integral and destructively at half-integral distances for $\Delta \phi = 0$, it will do exactly the opposite at $\Delta \phi = \pi$. Furthermore, addressing atoms transversally ($\Delta \phi = 0$) seems more favorable at typical magic wavelength trapping distances, e.g.\ $d/\lambda_0 \approx 0.58$ for $^{87}Sr$~\cite{campbell2008absolute,takamoto2005optical,ostermann2012cascaded}. Again, for $d \ll \lambda_0$ the mean field approach breaks down and one should rather turn to the Dicke model~\cite{dicke1954coherence}, reducing $N$ two-level emitters to one effective spin $N/2$-system~\cite{zhu2014quantum}.
Let us finally discuss the consequences for typical cases. Fig.~\ref{fig:timeevolution-chain} shows the time evolution of the average spin for an infinite chain initialized in a symmetric Ramsey state with either no phase shift or a phase shift of $\Delta \phi = \pi$ between neighboring atoms. The lattice constants have been chosen to be approximately $\lambda_0/2$ as would be typical~\cite{bloch2005ultracold}. We refrain from choosing exactly $\lambda_0/2$ to avoid the $1/\xi$ divergence. We observe that the dipoles' lifetimes vary strongly, comparing the subradiant behavior (red) where the collective dipole lives much longer than the natural lifetime of the atom to the superradiant (green) regime where the excitation vanishes very quickly. Additionally, to highlight the validity of the mean field approach, we add the results of a second order expansion simulation. Corresponding results for a full Ramsey sequence are shown in the appendix~\ref{appendix:ramsey-spectroscopy}.

{\it Conclusions --} In densely filled optical lattices dipole-dipole interaction and collective decay significantly change the evolution of an induced collective dipole. Due to the long range-nature of the coupling, sizable shifts appear even for long lived clock states despite their minute dipole moment, which limits accuracy and precision of Ramsey spectroscopy. Shifts and dephasing in large systems strongly depend on the dimensionality and geometry of the lattice, exhibiting resonant enhancements at particular lattice constants. While at current operating densities for Strontium~\cite{martin2013quantum, ushijima2014cryogenic, zhang2014spectroscopic} these shifts are smaller than other technical imperfections, they constitute inherent fundamental perturbations even in perfectly filled lattice clocks.

We have identified optimal operation geometries, which combine a negligible effective shift with a strong suppression of decay. In particular, for a 1D lattice with a tailored excitation angle and for a 2D hexagonal lattice favorable operation parameters for future generation clock setups were found. In this sense it seems possible to implement a high density dark exciton based atomic clock geometry with shifts many orders of magnitude below a single Hz and almost unlimited exciton life times. In 3D the interactions are particularly sensitive to the lattice constant and boundary effects, which dominate even for billions of particles rendering such setups very challenging.

While for most considerations we have focused on the case of clock transitions, the same physics is present in a more prominent and experimentally easier observable form for broader transitions. Optimizing geometries will also be relevant for devices such as superradiant lasers~\cite{bohnet2012steady, maier2014superradiant} or lattice based optical memories.

\begin{acknowledgments}
{\it Acknowledgments --} The authors acknowledge funding by the Austrian Science Fund FWF project SFB FoQus F4006-N13 (S.K., H.R.) and the DARPA QUASAR project (L.O., H.R.).
\end{acknowledgments}

\bibliographystyle{apsrev}

\onecolumngrid
\appendix

\section{Derivation of the mean field equations}
\label{appendix:meanfield-equations}

Starting from the full multiparticle density operator $\rho$ our master equation allows to obtain the following equations for the individual spin expectation values immediately
\begin{subequations}
\begin{align}
  \expect{\dot{\sigma_k^x}}  &=
      \sum_{j;j \neq k} \Omega_{kj} \expect{\sigma_j^y\sigma_k^z}
      -\frac{1}{2}\gamma \expect{\sigma_k^x}
        +\frac{1}{2}\sum_{j;j \neq k} \Gamma_{kj} \expect{\sigma_j^x \sigma_k^z}
\\
  \expect{\dot{\sigma_k^y}}  &=
      \sum_{j;j \neq k} \Omega_{kj} \expect{\sigma_j^x\sigma_k^z}
      -\frac{1}{2}\gamma \expect{\sigma_k^y}
     +\frac{1}{2}\sum_{j;j \neq k} \Gamma_{kj} \expect{\sigma_j^y \sigma_k^z}
\\
  \expect{\dot{\sigma_k^z}}  &=
      \sum_{j;j \neq k} \Omega_{kj} \Big(
                \expect{\sigma_j^x\sigma_k^y}
                - \expect{\sigma_j^y\sigma_k^x}
            \Big)
      -\gamma \big(1 + \expect{\sigma_k^z}\big)
      -\frac{1}{2}\sum_{j;j \neq k} \Gamma_{kj}\Big(
              \expect{\sigma_j^x \sigma_k^x}
              +\expect{\sigma_j^y \sigma_k^y}
            \Big).
  \label{eq:mf-timeevolution}
\end{align}
\end{subequations}
Assuming a spatially separable state $\rho = \bigotimes_k \rho_k$ leads to the lowest order mean field equations used in the letter.

\section{Mean field equations with tailored excitation phase}
\label{appendix:meanfield-phase-equations}

At zero temperature the ground state  $\rho = \bigotimes_k  \left( \ket{g} ( \bra{g}\right)_k $ is separable and in an idealized standard Ramsey procedure the first pulse would create a product state of equal weighted superpositions  $\rho = \bigotimes_k 1/2  \left( (\ket{g}+\ket{e}) (\bra{e}+ \bra{g})\right)_k $. This is the generic initial state we use in our work to study dipole-dipole interaction. In fact, this state possesses the maximal collective dipole moment and therefore shows strong interactions.

Of course, in any real setup this preparation step is not perfect as interactions are present during the excitation pulse and the excitation laser carries an intensity and phase gradient. Some of the errors can be corrected in improved excitation schemes\cite{yudin2010hyper,ostermann2013protected}.  However, particularly in extended systems a phase gradient is hard to avoid and will strongly influence the system dynamics. Fortunately, one can show, that a known phase gradient will not complicate the calculations too much.  If we allow for the individual atomic states to bare a spatially dependent phase of $\Delta \phi$ on the excited state, i.e. $\ket{\psi_k} = \frac{1}{\sqrt{2}} \left( \ket{g} + \exp (i \phi_k) \ket{e} \right)$,  we can absorb this into our equations efficiently. Using the abbreviations $\Omega_{kj}^\mathrm{cos} = \Omega_{kj} \cos(\phi_k - \phi_j)$ and $\Omega_{kj}^\mathrm{sin} = \Omega_{kj} \sin(\phi_k - \phi_j)$  we obtain the following modified equations of motion

\begin{subequations}
\begin{align}
  \frac{d}{dt}\expect{\tilde{\sigma}_k^x}
    &= \sum_{j;j \neq k} \Omega_{kj}^\mathrm{sin} \expect{\tilde{\sigma}_j^x\sigma_k^z}
            + \sum_{j;j \neq k} \Omega_{kj}^\mathrm{cos} \expect{\tilde{\sigma}_j^y\sigma_k^z}
        -\frac{1}{2} \gamma \expect{\tilde{\sigma}_k^x}
        +\frac{1}{2} \sum_{j;j \neq k} \Gamma_{kj}^\mathrm{cos} \expect{\tilde{\sigma}_j^x \sigma_k^z}
            -\frac{1}{2}\sum_{j;j \neq k} \Gamma_{kj}^\mathrm{sin} \expect{\tilde{\sigma}_j^y \sigma_k^z}
  \\
  \frac{d}{dt}\expect{\tilde{\sigma}_k^y}
    &= -\sum_{j;j \neq k} \Omega_{kj}^\mathrm{cos} \expect{\tilde{\sigma}_j^x\sigma_k^z}
            + \sum_{j;j \neq k} \Omega_{kj}^\mathrm{sin} \expect{\tilde{\sigma}_j^y\sigma_k^z}
        -\frac{1}{2} \gamma \expect{\tilde{\sigma}_k^y}
        +\frac{1}{2} \sum_{j;j \neq k} \Gamma_{kj}^\mathrm{sin} \expect{\tilde{\sigma}_j^x \sigma_k^z}
        +\frac{1}{2} \sum_{j;j \neq k} \Gamma_{kj}^\mathrm{cos} \expect{\tilde{\sigma}_j^y \sigma_k^z}
  \\
  \frac{d}{dt}\expect{\sigma_k^z}
    &= -\sum_{j;j \neq k} \Omega_{kj}^\mathrm{sin} (
                \expect{\tilde{\sigma}_j^x \tilde{\sigma}_k^x}
                + \expect{\tilde{\sigma}_j^y \tilde{\sigma}_k^y})
        +\sum_{j;j \neq k} \Omega_{kj}^\mathrm{cos} (
                \expect{\tilde{\sigma}_j^x \tilde{\sigma}_k^y}
                - \expect{\tilde{\sigma}_j^y \tilde{\sigma}_k^x})
      \nonumber\\&\qquad
        -\gamma (1+ \expect{\sigma_k^z})
        -\frac{1}{2} \sum_{j;j \neq k} \Gamma_{kj}^\mathrm{cos} (
                \expect{\tilde{\sigma}_j^x \tilde{\sigma}_k^x}
                + \expect{\tilde{\sigma}_j^y \tilde{\sigma}_k^y})
        -\frac{1}{2} \sum_{j;j \neq k} \Gamma_{kj}^\mathrm{sin} (
                \expect{\tilde{\sigma}_j^x \tilde{\sigma}_k^y}
                - \expect{\tilde{\sigma}_j^y \tilde{\sigma}_k^x}).
\end{align}
\end{subequations}
We see that the following definitions prove to be very helpful
\begin{subequations}
\begin{align}
  \Omega_k^\mathrm{cos} &= \sum_{j;j \neq k} \Omega_{kj} \cos(\phi_k-\phi_j)
  \qquad
  \Omega_k^\mathrm{sin} = \sum_{j;j \neq k} \Omega_{kj} \sin(\phi_k-\phi_j)
  \\
  \Gamma_k^\mathrm{cos} &= \sum_{j;j \neq k} \Gamma_{kj} \cos(\phi_k-\phi_j)
  \qquad
  \Gamma_k^\mathrm{sin} = \sum_{j;j \neq k} \Gamma_{kj} \sin(\phi_k-\phi_j)
\end{align}
\end{subequations}
Again, if we consider highly symmetric configurations where $\Omega^\mathrm{f} = \Omega^\mathrm{f}_k$ and $\Gamma^\mathrm{f} = \Gamma^\mathrm{f}_k$ and the rotated states are initially identical we can define the effective rotated quantities
\begin{align}
  \tilde{\Omega}^\mathrm{eff} &= \Omega^\mathrm{cos} - \frac{1}{2} \Gamma^\mathrm{sin}
  \\
  \tilde{\Gamma}^\mathrm{eff} &= \Gamma^\mathrm{cos} + 2 \Omega^\mathrm{sin}
\end{align}

which lead to a closed set of simplified effective equations as well, i.e.

\begin{subequations}
\begin{align}
  \frac{d}{dt}\expect{\tilde{\sigma}^x}  &=
      \tilde{\Omega}^{\mathrm{eff}}\expect{\tilde{\sigma}^y}\expect{\sigma^z}
      -\frac{1}{2} \gamma \expect{\tilde{\sigma}^x}
      +\frac{1}{2} \tilde{\Gamma}^{\mathrm{eff}} \expect{\tilde{\sigma}^x}\expect{\sigma^z}
  \\
  \frac{d}{dt}\expect{\tilde{\sigma}^y}  &=
      -\tilde{\Omega}^{\mathrm{eff}}\expect{\tilde{\sigma}^x}\expect{\sigma^z}
      -\frac{1}{2} \gamma \expect{\tilde{\sigma}^y}
      +\frac{1}{2} \tilde{\Gamma}^{\mathrm{eff}} \expect{\tilde{\sigma}^y}\expect{\sigma^z}
  \\
  \frac{d}{dt}\expect{\sigma^z}  &=
        -\gamma \big(1 + \expect{\sigma^z}\big)
        -\frac{1}{2} \tilde{\Gamma}^{\mathrm{eff}} \Big(\expect{\tilde{\sigma}^x}^2 + \expect{\tilde{\sigma}^y}^2\Big)
\end{align}
\end{subequations}

Note that such a phase gradient tends to mix the real and imaginary part of the interaction terms.

\section{Effective quantities for cubic lattices in 3D}
\label{appendix:effective-quantities-cubic}

In a cubic 3D lattice the number of neighbors at a given distance $r$ grows approximately as $r^2$.  Hence, one cam expect a slower convergence with distance. This problem is increased as the number of emitters to be considered grows with the third power of the system size. In contrast to 1D and 2D, together these two scalings prevent a convergence of the effective interaction parameters in the range of tractable lattices sizes of up to $N=(10^4)^3 = 10^{12}$ sites. Anyway, this is beyond experimentally realistic atom numbers so that we have to live with finite size effects.

\begin{figure}[ht]
  \center
  \includegraphics{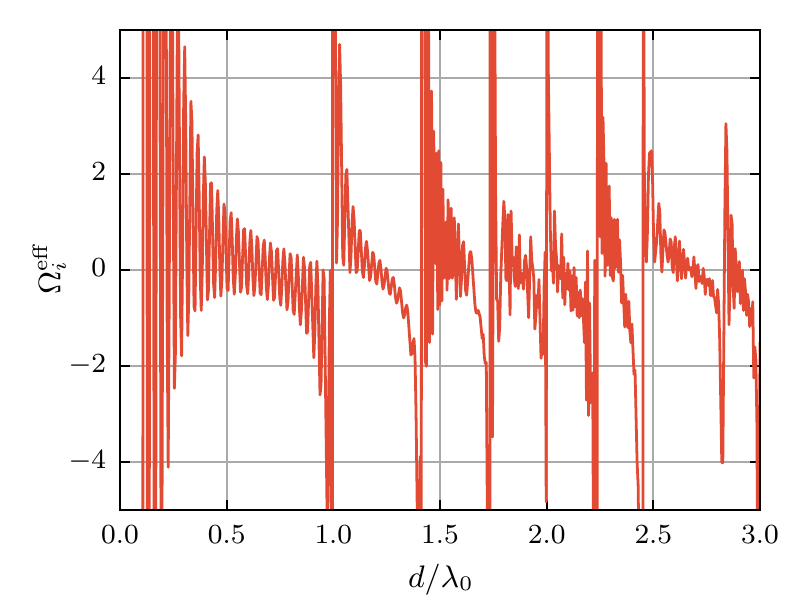}
  \includegraphics{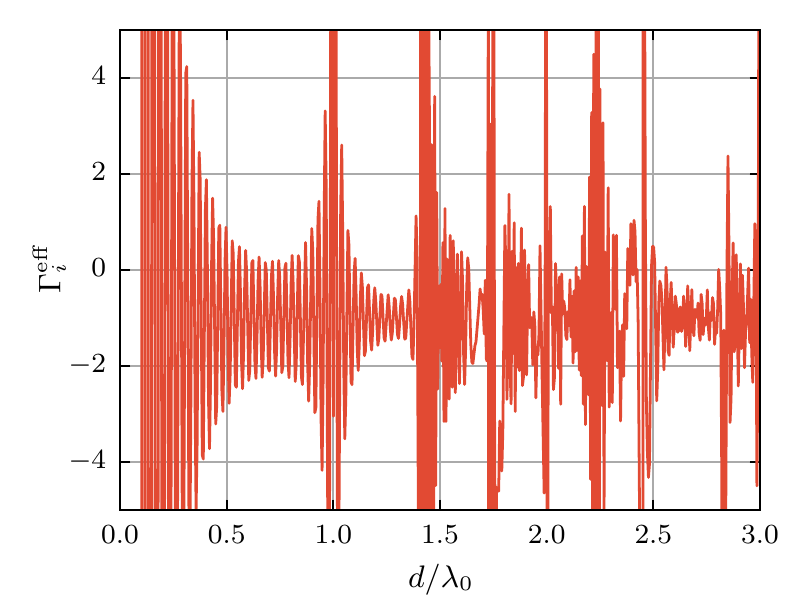}
  \caption{Effective quantities $\Omega^\mathrm{eff}_i$ and $\Gamma^\mathrm{eff}_i$ as experienced for the innermost spin inside a cube consisting of $2001 \times 2001 \times 2001$ spins in a cubic lattice configuration depending on the lattice spacing $d$. Even for very small changes of the lattice spacing the mean net-effect of all other spins will change dramatically.}
  \label{3d}
\end{figure}

In order to demonstrate the very slow convergence of the infinite range mean field model, we present some typical intermediate result for a 3D cubic lattice. In Fig. \ref{3d} we depict the effective coupling strengths $\Omega^\mathrm{eff}$ and $\Gamma^\mathrm{eff}$ for the innermost two-level system in a cubic lattice of about $8 \cdot 10^9$ particles, i.e. $2000$ particles in each direction. We obtain strong and very rapid oscillations of the shifts as a function of the lattice constant. Notice, that $1/r$-contributions as discussed in the letter will show up for planar and cubic diagonal distances of $\sqrt{2} \cdot r$ and $\sqrt{3} \cdot r$ as well. Increasing the atom number further still leads to changes of this pattern, so no final conclusions about physical properties and the behavior of a 3D cubic lattice can be obtained. However, perturbations of up to an order of magnitude larger than the linewidth as well as strong finite size shifts can be expected. In this case it is difficult to suggest an optimal lattice constant for a clock setup, except for avoiding certain resonances and choosing a region of about $d \approx 3\lambda/4$.

\section{Ramsey spectroscopy}
\label{appendix:ramsey-spectroscopy}

\begin{figure}[ht]
  \center
  \includegraphics{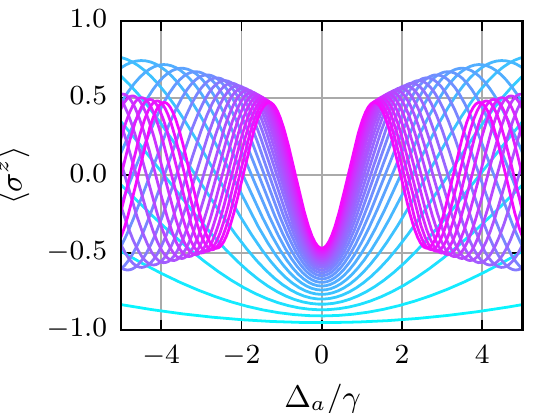}
  \includegraphics{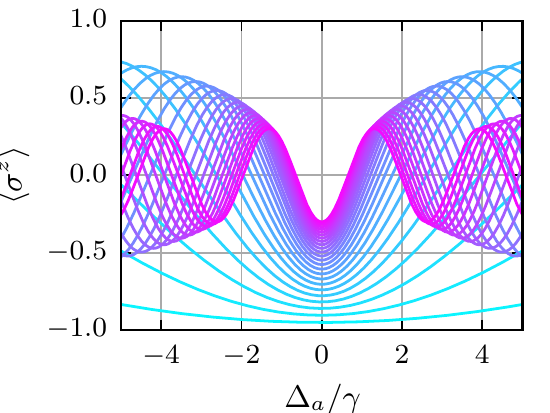}
  \includegraphics{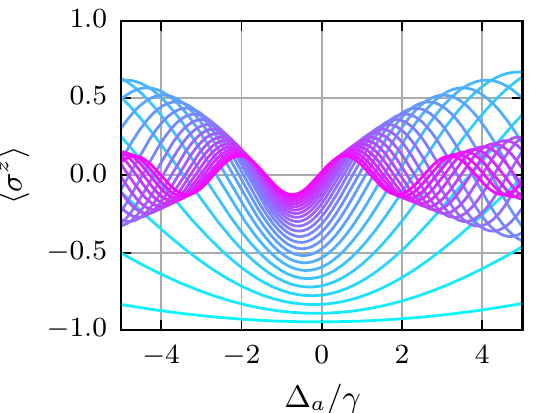}
  \caption{Simulated idealized Ramsey spectroscopy for different sets of effective cooperative interaction strengths, from left to right: subradiant case without shift $(\Omega^\mathrm{eff} = 0, \Gamma^\mathrm{eff} = -0.75)$, independent atom limit $(\Omega^\mathrm{eff} = 0, \Gamma^\mathrm{eff} = 0)$ and superradiant case with shift $(\Omega^\mathrm{eff} = 1, \Gamma^\mathrm{eff} = 1)$. The colors indicate the free evolution time with cyan representing a very short time and magenta meaning times up to $2.5\gamma^{-1}$.}
  \label{fig:ramsey-fringes}
\end{figure}

\begin{figure}[ht]
  \center
  \includegraphics{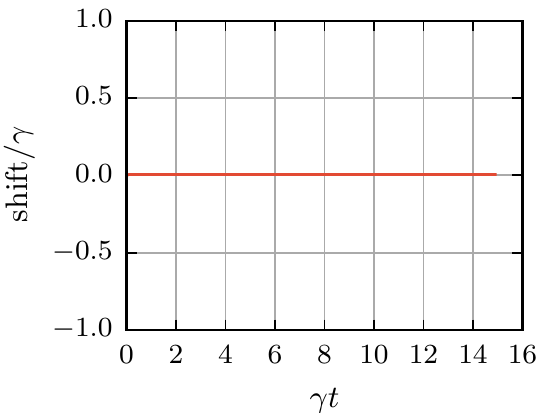}
  \includegraphics{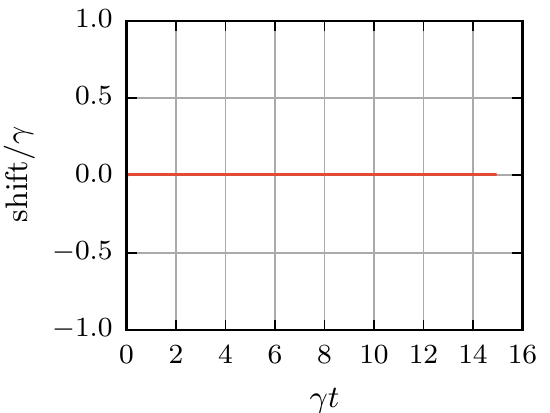}
  \includegraphics{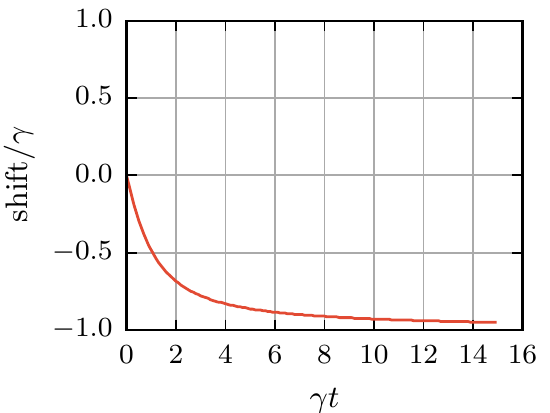}
  \caption{Shifts of the maxima after the free evolution time $\gamma t$ for different sets of effective interaction strengths as above, from left to right: $(\Omega^\mathrm{eff} = 0, \Gamma^\mathrm{eff} = -0.75)$, $(\Omega^\mathrm{eff} = 0, \Gamma^\mathrm{eff} = 0)$ and $(\Omega^\mathrm{eff} = 1, \Gamma^\mathrm{eff} = 1)$.}
  \label{fig:ramsey-shifts}
\end{figure}

\begin{figure}[ht]
  \center
  \includegraphics{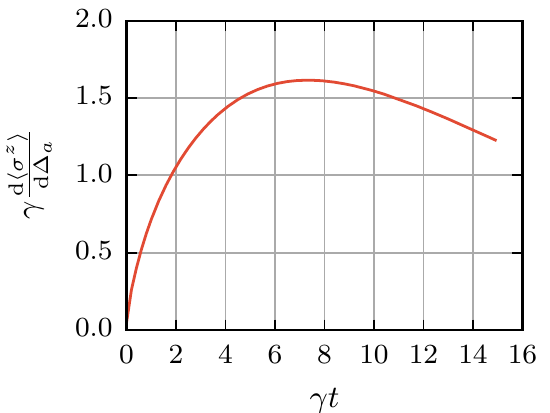}
  \includegraphics{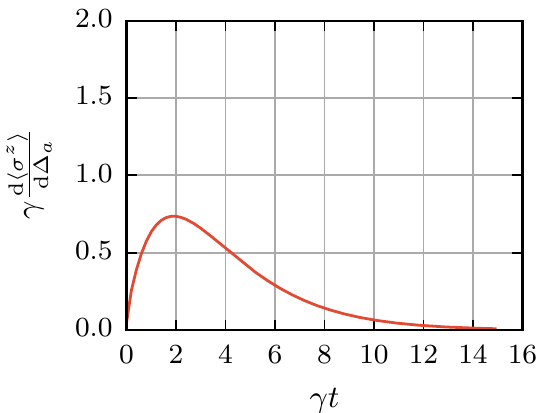}
  \includegraphics{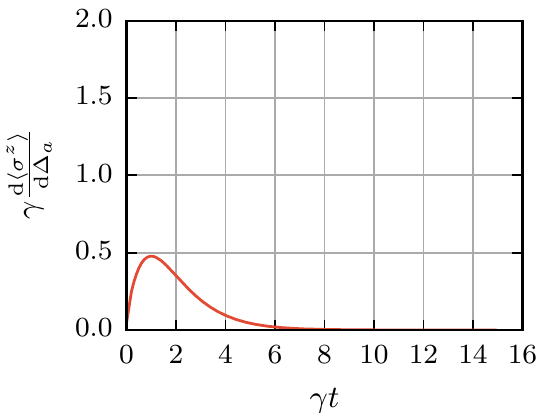}
  \caption{Slope of the signal at the zero crossing of the first fringe after a free evolution time of $\gamma t$ for different sets of effective interaction strengths as above, from left to right: $(\Omega^\mathrm{eff} = 0, \Gamma^\mathrm{eff} = -0.75)$, $(\Omega^\mathrm{eff} = 0, \Gamma^\mathrm{eff} = 0)$ and $(\Omega^\mathrm{eff} = 1, \Gamma^\mathrm{eff} = 1)$.}
  \label{fig:ramsey-derivatives}
\end{figure}

\begin{figure}[ht]
  \center
  \includegraphics{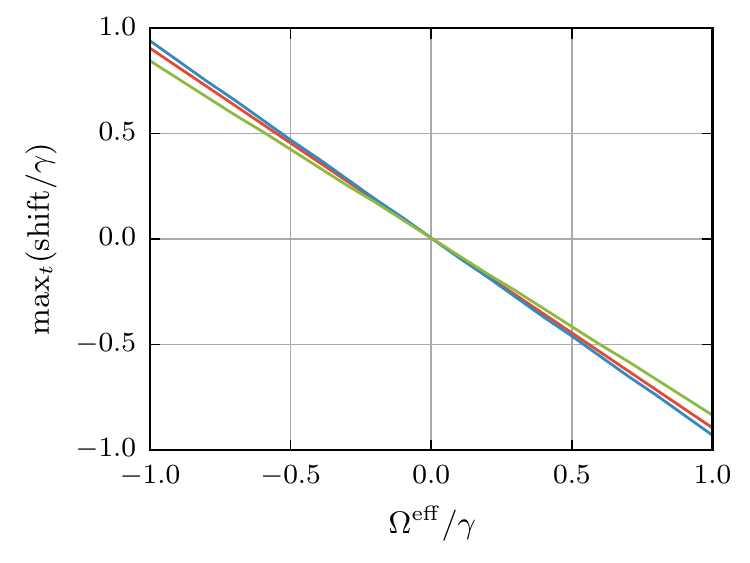}
  \includegraphics{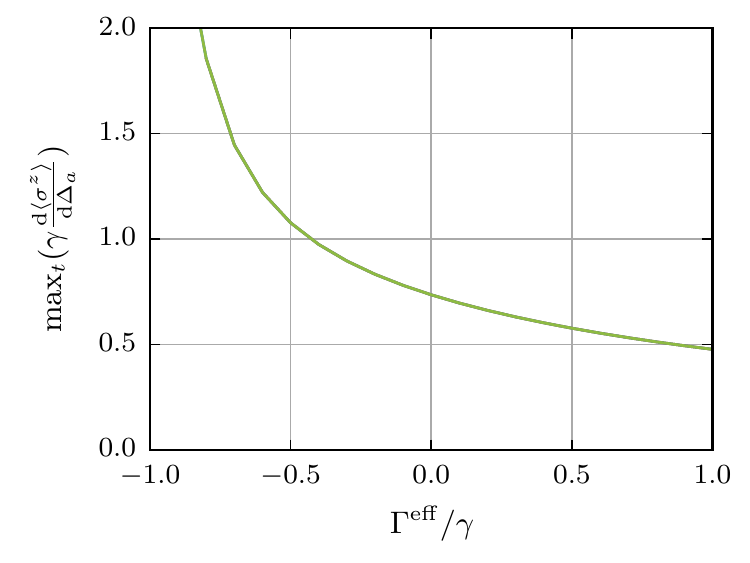}
  \caption{(a) Shift of Ramsey fringes depending on the effective coupling $\Omega^\mathrm{eff}$ after $t=15\gamma^{-1}$}. The different lines represent different choices of $\Gamma^\mathrm{eff}$, which hardly influence the result. The fringe shifts follow the effective mean field dipole coupling $\Omega^\mathrm{eff} $ almost linearly and thus can be read off from the figures in the main manuscript. (b) Maximally achievable slope at roots depending on $\Gamma^\mathrm{eff}$. The result is independent of the choice of $\Omega^\mathrm{eff}$. Note that a negative $  {\Gamma}^\mathrm{eff} $ improves the measurement precision beyond the independent atom value.
  \label{fig:ramsey-maximums}
\end{figure}

The effective coupling and decay parameters  $\Omega^\mathrm{eff}, \Gamma^\mathrm{eff}$ characterize the interaction induced perturbation of the individual spin dynamics. Consequently, they will alter the Ramsey signal by introducing shifts of the fringes and modifications of the maximally obtainable contrast. As the actual connection between the magnitude of these effective couplings and their quantitative effect on the signal is nontrivial, we demonstrate the alterations of the Ramsey signal in the following examples. Using the previously derived equations of motion, it is straight forward to simulate the results of an ideal Ramsey sequence. By starting with a $\pi/2$-pulse all spins are rotated into the x-direction of the Bloch sphere. For a time $\gamma t$ the system evolves according to the equations
\begin{subequations}
\begin{align}
  \expect{\dot{\sigma^x}}  &=
      -\Delta_a \expect{\sigma_k^y}
      +\Omega^{\mathrm{eff}}\expect{\sigma^y}\expect{\sigma^z}
      -\frac{1}{2} \gamma \expect{\sigma^x}
      +\frac{1}{2} \Gamma^{\mathrm{eff}} \expect{\sigma^x}\expect{\sigma^z}
  \\
  \expect{\dot{\sigma^y}}  &=
      \Delta_a \expect{\sigma_k^x}
      -\Omega^{\mathrm{eff}}\expect{\sigma^x}\expect{\sigma^z}
      -\frac{1}{2} \gamma \expect{\sigma^y}
      +\frac{1}{2} \Gamma^{\mathrm{eff}} \expect{\sigma^y}\expect{\sigma^z}
  \\
  \expect{\dot{\sigma^z}}  &=
        -\gamma \big(1 + \expect{\sigma^z}\big)
        -\frac{1}{2} \Gamma^{\mathrm{eff}} \Big(\expect{\sigma^x}^2 + \expect{\sigma^y}^2\Big)
\end{align}
\end{subequations}
where $\Delta_a = \omega_0 - \omega_L$ is the detuning between the probe laser and the atomic transition frequency. After this free evolution a second $\pi/2$-pulse is applied and, finally, the expectation value of $\sigma^z$ is measured. For a given system characterized by the effective quantities $\Omega^\mathrm{eff}$ and $\Gamma^\mathrm{eff}$ the result of this measurement depends on the waiting time as well as on the detuning $\Delta_a$. In Fig.~\ref{fig:ramsey-fringes} the outcome of this numerical experiment is shown for three different realistic sets of effective quantities. The decisive quantity for the accuracy with regards to atomic clocks is the shift of the fringes due to the dipole-dipole interaction which can be obtained by measuring the shift of the maxima of the Ramsey fringes. The shifts for the chosen examples are shown in Fig.~\ref{fig:ramsey-shifts}. On the other hand the slope of the fringes at their roots is the determining factor for the best achievable experimental precision. The numerical results are shown in Fig.~\ref{fig:ramsey-derivatives}. As seen in Fig.~\ref{fig:ramsey-maximums} the maximal shifts depend on $\Omega^\mathrm{eff}$ only, while the maximal slope at zero points is governed by $\Gamma^\mathrm{eff}$. For realistic values for the effective quantities this means the accuracy can be limited to $\gamma$ and the achievable precision can vary by a factor of $5$.

\end{document}